\documentclass[]{aastex631}

\usepackage{url}
\begin{document}

\title{The Relationship between the Kinematics of Coronal Mass Ejections and the Brightness of the Corona}

\correspondingauthor{Kelly Victor-French}
\email{kelly.d.victor-french.civ@us.navy.mil}

\author[0000-0003-4070-5518]{Kelly Victor-French}
\affiliation{U.S. Naval Research Laboratory \\
4555 Overlook Ave SW \\
Washington, DC 20375, USA}

\author[0000-0002-8692-6925]{Karl Battams}
\affiliation{U.S. Naval Research Laboratory \\
4555 Overlook Ave SW \\
Washington, DC 20375, USA}

\author[0000-0002-4998-0893]{Brian E. Wood}
\affiliation{U.S. Naval Research Laboratory \\
4555 Overlook Ave SW \\
Washington, DC 20375, USA}

\begin{abstract}

We present an investigation into an apparent relationship between white-light coronal brightness and the kinematics of flare-associated CMEs. Using a unique dataset known as the LASCO Coronal Brightness Index (CBI), we conduct a study that explores the brightness in the lower solar corona and its relationship to the velocity of flare-associated coronal mass ejections (CMEs). We analyze all M- and X-class flares that take place on or near the limbs of the Earth-facing disk of the Sun between 1996 and 2022, determine if these flares are associated with CMEs, and record the projection-corrected velocity of the eruptions if they occurred. Using the CBI dataset, we evaluate the brightness in the corona directly overlying the flare source locations between 2.4 and 6.2 solar radii, and find that above a certain level of coronal brightness, the likelihood of a high-velocity CME significantly decreases. This result implies coronal brightness could be an important indicator of the kinematics of solar CMEs. We also highlight and discuss the unique nature of Active Region (AR) 12192 in 2014, observing that its unprecedented overlying coronal brightness may be related to the low CME productivity of that region.

\end{abstract}


\section{Introduction} \label{sec:intro}

Space weather describes the conditions in and around Earth's magnetosphere, with the space weather environment largely driven by events that originate on the Sun. Advanced knowledge of the processes that occur on the Sun and how they will affect technology and humans in space, and life on Earth, is at the core of effective space weather forecasting. The Sun may be many millions of miles away from Earth, but it only takes a matter of minutes, hours, or days for Earth to experience the effects from the variety of space weather related events. This relatively short timescale for the arrival of space weather effects makes early warning indicators imperative.

Unfortunately, the Sun and Earth's magnetosphere are complex systems, and therefore the space weather environment is difficult to forecast. Coronal mass ejections (CMEs) are one of the most important space weather events. They can become hazardous to space missions and disruptive to life on Earth, but their effects are challenging to predict due to variable parameters such as the eruption's direction, magnetic field orientation, and velocity. At Earth, CMEs can become drivers of geomagnetic storms which can damage satellites and even cause ground-induced currents, both of which can lead to interruption or failure of critical navigation and communication systems upon which societies rely \citep{Baker04}.

CMEs are known to be associated with other solar phenomena such as prominence eruptions and solar flares, \citep[e.g.,][]{Webb12}, with most solar flares originating from solar active regions (ARs). Both solar flares and CMEs are difficult to predict, and space weather forecasters generally do not know if one or the other will occur until it has already happened and is detected by a telescope \citep{Koskinen17}. We only know about a flare after it is observed near the erupting region on the Sun by an EUV or X-ray telescope, while in the case of a CME, it is unlikely that we will know whether one will occur until it has been visually detected in coronagraph imagery from a few solar radii away from the Sun. Coronagraphs are therefore important to space weather forecasting and monitoring, as they are the primary visual indicator of CMEs. This coronagraph imagery also enables forecasters to estimate the direction, velocity, and mass of CMEs, all of which can be fed into models to predict their potential geoeffectiveness. While other instruments, such as EUV imagers \citep{Bewsher08, Mason16}, can often indicate CMEs and some of their properties, our study focuses on coronagraphs.

The most widely used coronagraph of the past two decades has been the Large Angle and Spectrometric Coronagraph Experiment \citep[LASCO,][]{Brueckner95} on the joint ESA/NASA Solar and Heliospheric Observatory \citep[SOHO,][]{Domingo95}. SOHO is in a halo orbit around the first Lagrange point (L1), which is a region of stability between Earth and the Sun. This location gives LASCO an uninterrupted view of the Earth-facing side of the Sun, which has enabled early visual detection of thousands of CMEs \citep{Vourlidas10}, including many Earth-directed events. Other important coronagraphs to space weather forecasting are the COR1 and COR2 constituents of the Sun Earth Connection Coronal and Heliospheric Investigation \citep[SECCHI][]{Howard08} package of imagers on the Solar Terrestrial Relations Observatory \citep[STEREO][]{Kaiser08}. This mission has provided additional views of the solar corona, which can be valuable. We touch upon this later in Section~\ref{sec:discuss}.

While it is no surprise that the heliophysics community has largely focused on eruptive events, it is also of great interest to study events that would have been expected to occur but did not. For example, the anomalous AR12192 in Fall of 2014, has been investigated by numerous authors for its size, complexity, and frequency of powerful flares \citep{Sun15, Chen15}. It is noteworthy that this region produced few CMEs, and certainly none of any note \citep{Sun15, Sarkar18}. Understanding the mechanisms driving the behavior of regions like AR12192 could yield valuable information about CME kinematics. 

Previous studies of event eruptivity have largely remained focused at the event origin on or below the solar disk, and the magnetic conditions surrounding the (non-)eruption. \cite{Chen15} explored non-eruptive flares from AR12192 using the Atmospheric Imaging Assembly (AIA) and Helioseismic and Magnetic Imager (HMI) aboard the Solar Dynamics Observatory (SDO) and found evidence of strong confinement from the overlying magnetic field. They suggest this could be responsible for the absence of CMEs from the region. \cite{Liu16} also used data from SDO/HMI to explore the size and complexity of this AR. Their results suggest that large ARs will always be flare-productive, but whether or not they produce CMEs, is dependent on the presence of a mature sheared core, and weak overlying arcades. \cite{Thalmann19} noted that the ratio of current-carrying to total magnetic helicity can be a strong indicator of the eruptive potential of an AR, but is not a sufficient indicator for major confined flares, like those seen in AR12192. These are just a few of the studies that have been done on this AR which, due to its peculiar nature, has drawn particular attention from the solar physics community.

Here, we take an alternative approach to investigate event kinematics by looking ``up'' at the white-light corona overlying solar flare source locations, which are often associated with ARs. With LASCO observations spanning over 29 years and two complete solar cycles, there exists a significant dataset for studying the long-term behavior of the corona and the coronal response to the solar cycle. This is well illustrated by the LASCO-based Coronal Brightness Index (CBI) \citep{Battams20} – a reduction of the entire LASCO C2 archive to a single data product that captures the global brightness of the solar corona in a reduced data format, highlighting the modulation of the brightness of the K (electron) corona over the past two solar cycles \citep[e.g., Figure 3b of ][]{Battams20}. In addition to showing the global brightness of the corona, CBI can be used to study relatively small, localized regions in the corona. This data allows us to consider the question of whether overlying coronal white-light brightness affects, or is at least related to, the eruptiveness of flaring source locations and the velocity of any associated CMEs. 

In this study, we use CBI to explore if there is a relationship between coronal brightness and CME occurrence and kinematics. A demonstrable link between CBI values and CME occurrence and velocity would potentially be of value to space weather forecasting. Knowledge of whether a solar flare source location is likely to produce many fast CMEs would help to provide a clearer picture of the expected future space weather environment. The space weather implications from this current study should be further explored and addressed in future work. In this paper, we begin with an overview of the LASCO mission and the CBI dataset. Next, we describe the catalog we used for data collection and the steps we took to prepare the data to meet criteria for our research. In the following section, we explain how we analyzed the localized corona above flaring source locations. Finally, we discuss our results and present conclusions.

\begin{figure}[ht!]
\plotone{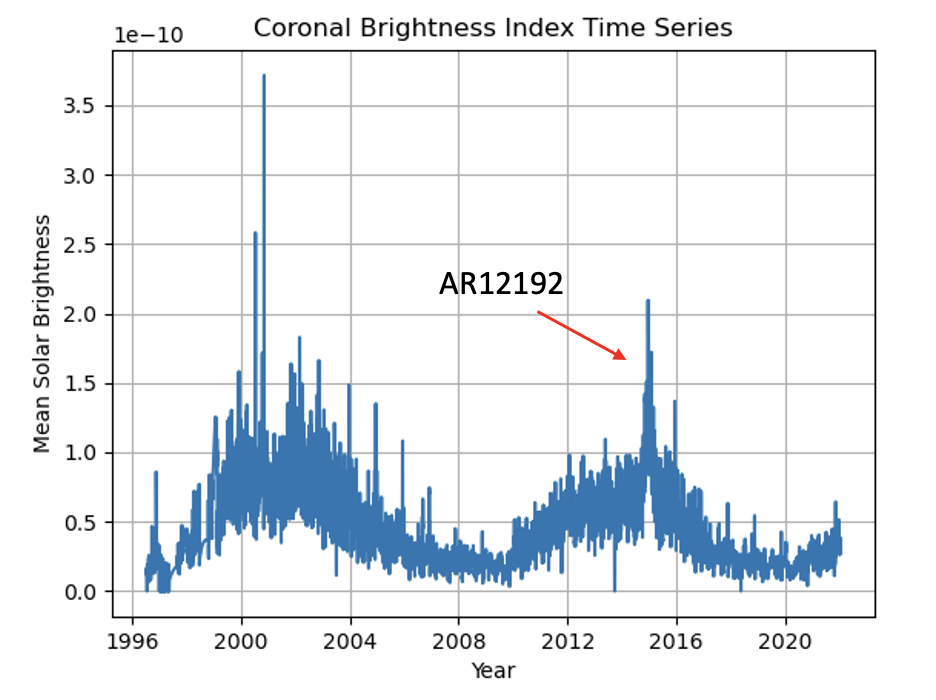}
\caption{CBI time series of the mean solar brightness of the corona as seen in LASCO C2 from April 1996 to April 2022. Each data point is a daily median of the mean solar brightness, in other words a global median CBI value. In October 2014, AR12192 stands out as unusual, sustained brightness. All individual outlier spikes are short-term effects, primarily from bright planet transits.}
\label{fig:timeseries}
\end{figure}

\section{Data Collection and Criteria} \label{sec:data}

In this study, we used data from SOHO/LASCO \citep{Brueckner95, Domingo95} and the Atmospheric Imaging Assembly \citep[AIA,][]{Smith06} on the Solar Dynamics Observatory \citep[SDO,][]{Pesnell12}. Information on flares and CMEs was collected from the Coordinated Data Analysis Workshop (CDAW) website\footnote{\url{https://cdaw.gsfc.nasa.gov/pub/yashiro/flare_cme/fclist_pub.txt}}. In the following Sections, we will describe these missions and the data products central to this investigation. 

\begin{figure}[ht!]
\plotone{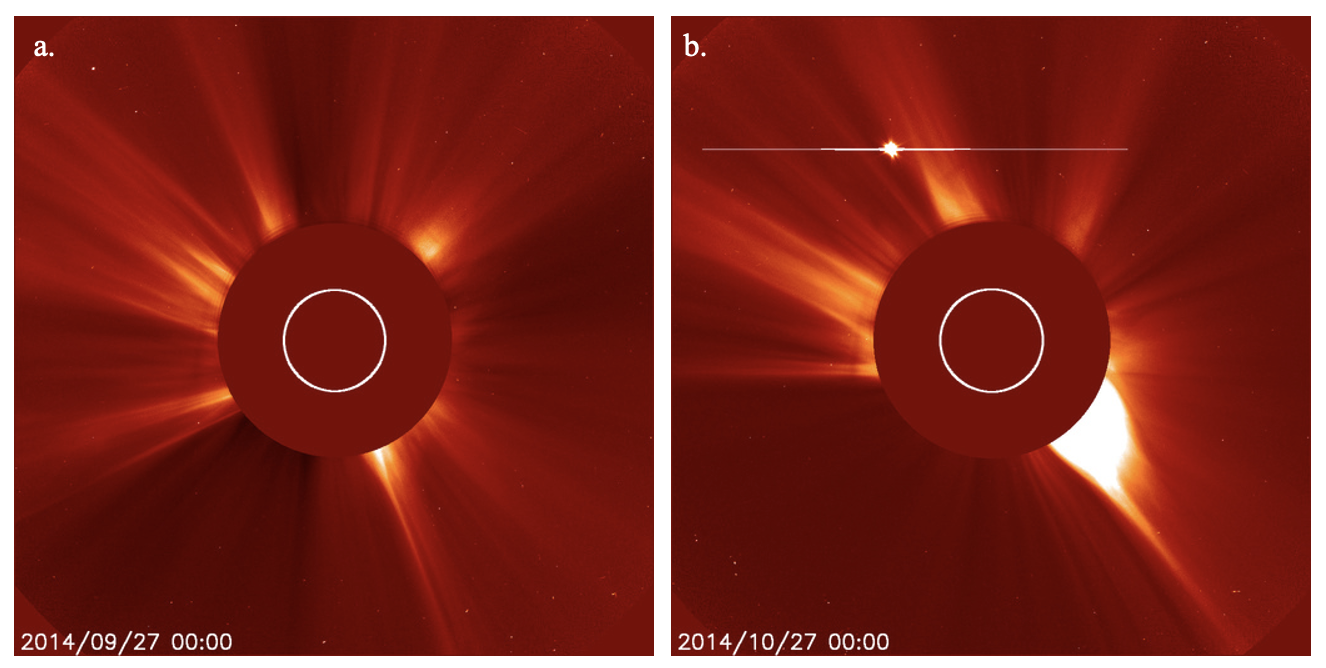}
\caption{(a) An image from LASCO C2 when AR12192 was not visible in the corona. (b) October 27, 2014, when AR12192 was near the west limb of the sun and the overlying corona was anomalously bright in C2 coronagraph imagery, per Figure~\ref{fig:timeseries}.}
\label{fig:bright_ar}
\end{figure}

This study was motivated by an observation first made by \cite{Wang18} and again by \cite{Battams20}, who noted a substantial spike in coronal brightness in 2014 coinciding with AR12192. This is represented here in Figure~\ref{fig:timeseries} which shows a time series of mean coronal brightness in CBI observations from 1996 to 2022, and identifies an anomalous sustained brightness peak corresponding to the appearance of AR12192 \citep{Battams20}, the sharper peaks in the figure being artifacts of planet transits. This active region produced numerous X-class flares, but no CMEs over 500 kms$^{-1}$ \citep{Sun15}. Figure~\ref{fig:bright_ar}(a) shows a nominal LASCO C2 image on September 27, 2014, and one month later (b), shows the extremely bright corona that accompanied this AR. With this peculiar behavior, and also having the brightest observed white-light coronal presence observed in SOHO's 29 years, it raises a question of whether there could be a link between that exceptional coronal brightness and the lack of high-velocity events from that AR. More broadly, we can consider a larger-scale question of whether coronal brightness affects, or is at least related to, the kinematics of any resulting CMEs from flaring source locations.

\subsection{LASCO C2 and the Coronal Brightness Index} \label{subsec:Coronal Brightness Index}

SOHO was launched in December 1995 and began routine operations in April 1996 from its vantage point in a halo orbit around the L1 Lagrange point. LASCO is one of eleven instruments on board the SOHO spacecraft, consisting of three coronagraphs, known as C1, C2, and C3, that record images of the Sun's corona from 1.1 to 30 solar radii ($R_\Sun$). Our study focuses on data from the C2 instrument, which records Sun-centered white-light observations from 1.5 to 6.2$R_\Sun$. Nominal LASCO C2 observations are 1024x1024-pixel images, with an 11.9 arc-seconds per pixel resolution. Between mission launch and August 2010, observations were generally recorded at a cadence of three full-resolution observations per hour, plus periodic sets of lower resolution, filtered calibration images. After a change in operations status in 2010, the cadence was increased to five full-resolution images per hour. The LASCO C1 telescope stopped functioning in 1998 following a spacecraft anomaly, and the C3 camera does not currently have a CBI product, so data from these coronagraphs are not included in this study. 

This study relies on the LASCO Coronal Brightness Index (CBI) \citep{Battams20}, which is a reduction of the entire C2 archive from 1996 through 2022. For each day of nominal observations, all fully calibrated images were reduced into a single daily median image, with physical units of mean solar brightness (MSB). For each of these daily observations, mean daily brightness values were extracted from 0.1$R_\Sun$ $\times$ 1\textdegree~regions for a full 360\textdegree~to produce a single CBI data array (a ``CBI slice'') at a given height above the solar limb. Each daily median contains arrays that were extracted at 38 different heights in the corona, from 2.4 -- 6.2$R_\Sun$, to produce the final data product, resulting in a data cube that is 360~(degrees)~$\times$~38~($R_\Sun$)~$\times$~9397~(days). For this study, we extracted daily medians at all heights.

\begin{figure}[ht!]
\plotone{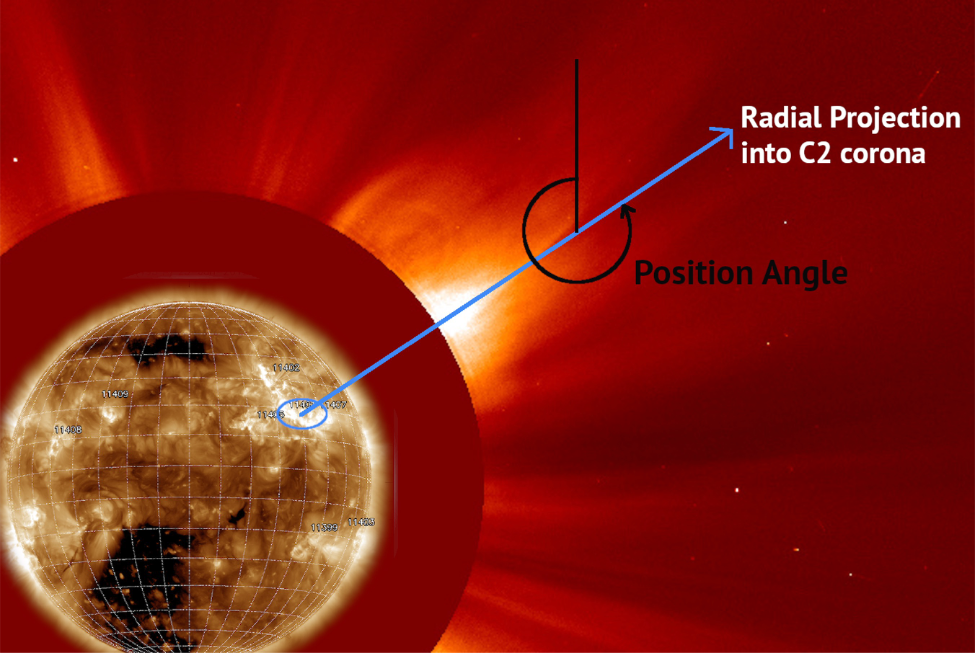}
\caption{AR11401 on the solar disk as seen by SDO AIA 193 Angstrom on January 24, 2012 at 13:00UT, the bright corona directly above it, and a radial projection from the AR into the corona as seen by LASCO C2.}
\label{fig:radproj}
\end{figure}

The premise of this study was to determine quantitative CBI-based values for the brightness of the corona directly overlying flaring source regions and look for any apparent relationships between these values and CME kinematics. Due to the C2 occulting disk, most of the corona that is visible is associated with near-limb activity. Accordingly, we only included events that are 45\textdegree~in longitude or less from the limb. Throughout this study, we will refer to this location as the limb, or near-limb, and refer to such events as ``limb events''. Figure~\ref{fig:radproj} shows a portion of a nominal LASCO C2 image from January 24, 2012, in which we observe the bright corona above a flare source location on the disk of the Sun. Superimposed on this, we can see AR11401 (circled), which was associated with the flare source location on the disk of the Sun as shown in the corresponding SDO AIA in 193 Angstroms. The blue line in the image is an illustration of a radial projection from the Sun's surface into the corona, originating at the cataloged source location. This is calculated by projecting a radial vector from a source region into a position angle on the LASCO plane of sky, based upon converting the heliographic coordinates to heliocentric-cartesian, and then heliocentric-radial \citep{Thompson06}. We then use the value of this position angle to extract the relevant portions of the CBI dataset and evaluate the brightness above that particular source location. The uncertainties associated with such an assumption are noted in Section~\ref{sec:discuss}. 

\begin{figure}[ht!]
\plotone{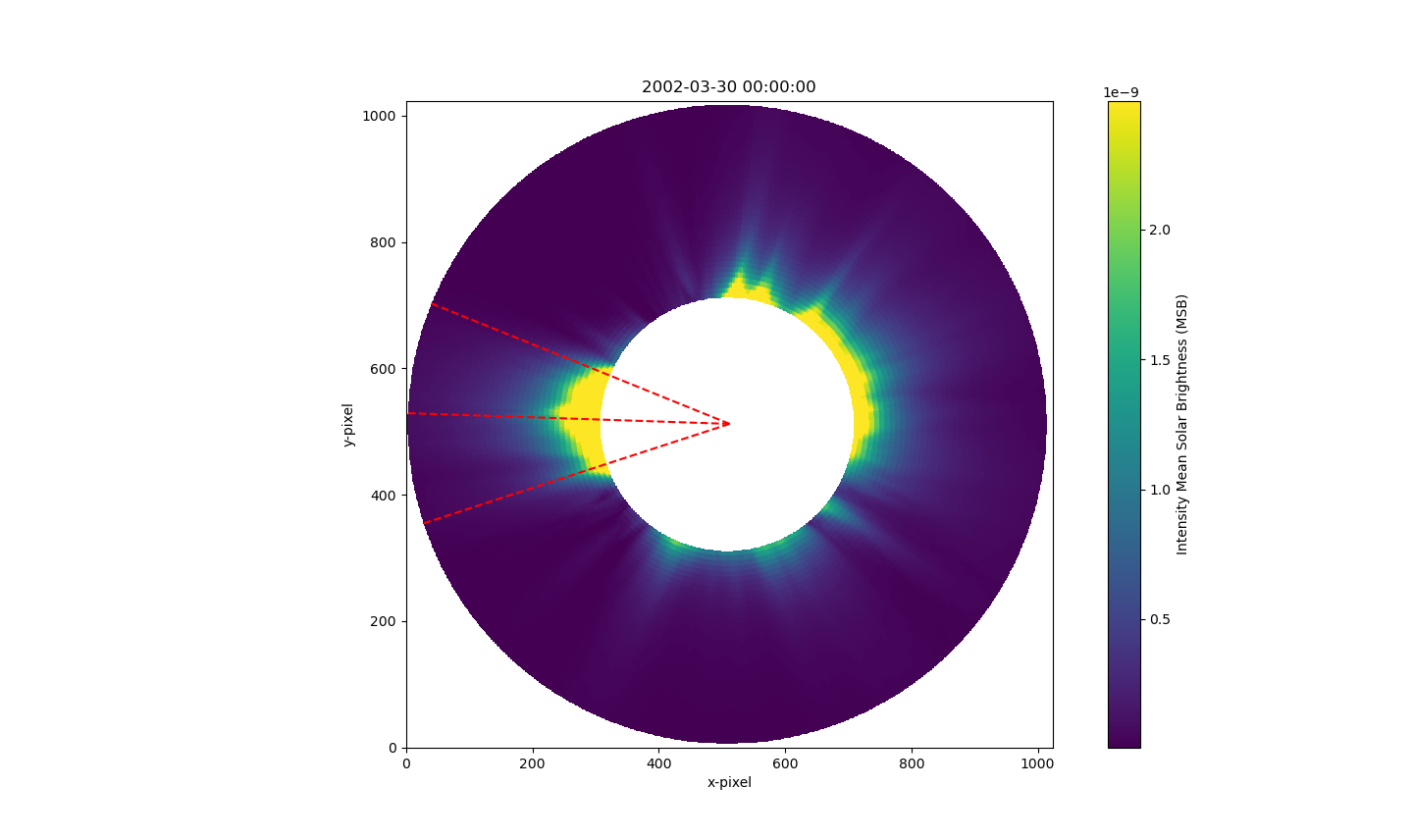}
\caption{CBI slice for March 30, 2002. This specific date is arbitrary, but this is a standard example of what an individual day of data from the CBI cube looks like. The central dashed red line is a radial projection at position angle 88\textdegree~that has been drawn out into the corona. A 40\textdegree~wedge is centered around the 88\textdegree~position angle. The x- and y- here correspond to the horizontal and vertical axes of a LASCO image. A color bar shows the intensity of the coronal brightness in MSB.}
\label{fig:slicewedge}
\end{figure}

An example of a single day of data, a ``CBI slice'' is shown in Figure~\ref{fig:slicewedge}, where the $360 \times 38 \, R_\odot$ array has been projected back into an x,y plane similar to the LASCO field of view, where x is the horizontal axis in the LASCO-like image, and y is the vertical axis. The central dashed line shows a radial projection at 88\textdegree, and the outer lines demonstrate a 40\textdegree~wedge-shaped region of interest (hereafter referred to as ``wedges'') centered around the position angle. The position angle and wedge width correspond to the source region of an example event discussed further in Section~\ref{sec:analysis}. Each slice of CBI data is the median of daily observations made by LASCO C2, with an appropriate background model used to subtract out the effect of the dominant F-corona signal, as described in \cite{Battams20}. The advantage of using a median of daily observations is that all transient features, such as CMEs, stars, and particle storms, are mostly removed, leaving behind more stable structures such as streamers and white-light coronal brightness above flare source locations. 

As indicated in Figure~\ref{fig:radproj}, SDO/AIA data was used to help associate source locations on the solar disk with the overlying corona. Typically, this is a straightforward process, as a line is projected radially from the flare source location on the solar disk (as listed in the CDAW data) out to the corona, but occasionally we find that the brightness in the corona is slightly shifted from the flare source. This offset could be caused by a number of factors, including the presence of coronal holes near the source location, which could only be seen using SDO/AIA imagery.

\subsection{CME/flare observations and data preparation} \label{subsec:CME/flare observations and data preparation}

CME and solar flare data were collected from the SOHO LASCO CME Catalog from the Coordinated Data Analysis Workshops (CDAW) Data Center \citep{Yashiro04}, which lists all recorded CMEs from the LASCO archive. Each entry contains a height-time file, which is a text file that lists solar radii height measurements of a CME as a function of time. The file also records start time, linear velocity, position angle, and other parameters that describe the CME\footnote{\url{https://cdaw.gsfc.nasa.gov/pub/yashiro/flare_cme/fclist_pub.txt}}. Data collection began with searching for all M- and X-class solar flares and their associated CMEs that occurred within 45\textdegree~of the solar limb. For each associated CME, the start time, linear velocity, and flare source location from the height-time files, were recorded. The time range for the events analyzed followed the time range available in the CBI dataset, 1996 -- 2022. For this analysis, we could only include limb or near-limb events, since the brightness of the overlying corona is not observable by LASCO for disk center events. Accordingly, we only included events that were 45\textdegree~in longitude or less from the limb. For example, an event with the source location N05W78 would meet the criteria for inclusion, but an event with the source location N05W44 would not. The choice of 45\textdegree~is somewhat arbitrary, but sufficiently excludes center disk events which would be entirely blocked by the occulter and have no limb brightness. The goal was to associate flare source locations with their overlying coronal brightness as observed in the CBI data. It is also worth noting that while many of the events in our dataset were associated with ARs, some source locations were not assigned named ARs. Moreover, there was sometimes ambiguity in identifying the flare-CME relationship. We did not make any attempts to adjust those associations in the CDAW database, and instead omitted some events for which ambiguous, insufficient, or incomplete records were available. 

\section{Analysis} \label{sec:analysis}

We conducted a systematic study of all flaring events that fit the criteria described in Section~\ref{sec:data}. This did not discriminate between eruptive and non-eruptive events -- both were included and necessary for determining if coronal brightness could play a role in CME suppression. 

Each flare source location, as identified in the original CDAW database, was converted to a position angle in the LASCO C2 image by assuming a 3D radial projection from the disk onto the 2D plane of sky. We analyzed each region of localized coronal brightness by extracting an angular wedge of identical size for each event from the appropriate portion of CBI data (see Figure~\ref{fig:slicewedge}, for an example). We explored several different angular wedge sizes and eventually settled on a 40{\textdegree} wedge angle, 20{\textdegree} above and below the position angle. Although ARs and flaring source locations, and ergo their overlying coronal brightness, are not of identical sizes, 40{\textdegree} presented a reasonable approximation for the majority of sources. The choice of 40{\textdegree} is admittedly somewhat subjective. Future studies may benefit from more sophisticated procedures for choosing wedge sizes for specific events.

Throughout the analysis, we found that some position angles did not perfectly capture the bright region in the corona overlying the flare source location. This could be due to deflection from a feature on the Sun, such as a coronal hole, or a listed flare source location that was slightly off due to the inherent ambiguity of assigning source locations to solar events. For this reason, we manually (by-eye) adjusted the position angle of some events to be better centralized on the apparent overlying corona. Out of 1,264 events, we adjusted 411 position angles, or approximately 33\%. Adjustments were usually small, with a mean adjustment of 8.38{\textdegree}. This dataset, which contains all events that fit our criteria described in Section~\ref{sec:data}, is sorted from brightest to least bright (CBI) event and contains the date, flare intensity, source location, summed CBI wedge value, and CME velocity. 

Due to projection effects, CME velocities from single spacecraft observations are often unreliable \citep{Paouris21}. To account for this, we applied the following equation from \cite{Paouris21} to the apparent heights listed in the height-time files for each CME entry to ``correct'' them for this projection effect. 
\begin{equation}
h = r \sqrt{1 - \cos^2(\phi) \cos^2(\theta)}
\label{eqn_rsun_adj}
\end{equation}
Here, $\phi$ and $\theta$ are the latitude and longitude values of the source region on the solar disk, and \textit{r} is the apparent (plane of sky) solar radius height at which each measurement was taken. Applying Equation~\ref{eqn_rsun_adj} provided a new ``true'' solar radius height, \textit{h}, which we then used to calculate a new corrected velocity for each CME. These corrections are generally modest, given that our events are chosen to be closer to the limb. The average value of these corrections is 28.4 km/s. After this data preparation, we were left with 1,264 events for analysis.

\section{Results} \label{sec:res}

\begin{figure}[ht!]
\plotone{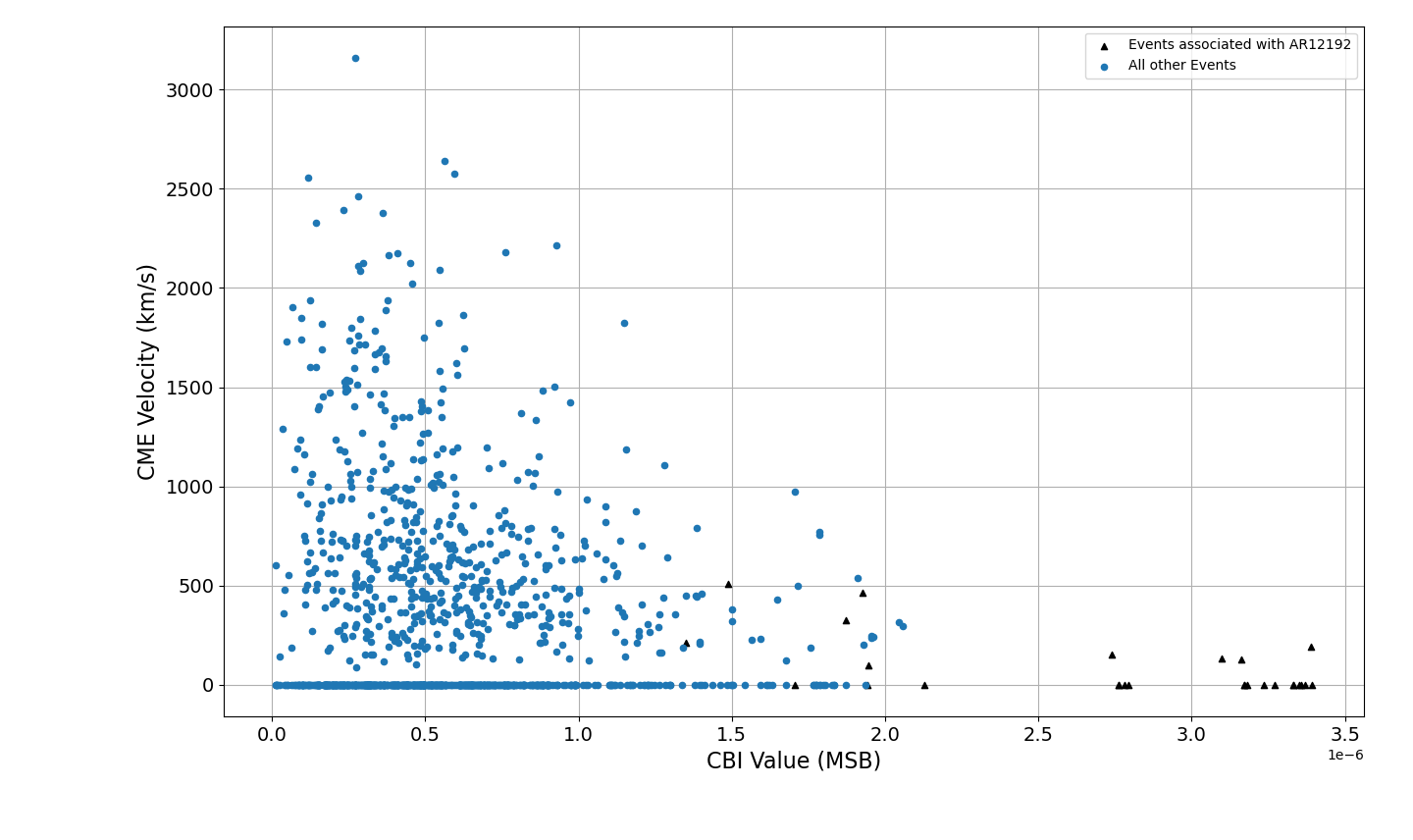}
\caption{CME Velocity in kms$^{-1}$ as a function of the CBI value in mean solar brightness. Each CBI value is the summed brightness from each 40{\textdegree} wedge. The small black triangles mark events associated with AR12192. The blue circles represent all other events and their associated velocities. The zeros correspond with flares that had no associated CMEs}.
\label{fig:results}
\end{figure}

 We applied the stated methodology of extracting 40{\textdegree} wedges in the overlying corona and recorded the summed brightness for each event. In Figure~\ref{fig:results}, we present our results as a scatter plot, showing the CME velocity in kms$^{-1}$ of each event versus the CBI value in a 40{\textdegree}wedge, with units of MSB. We observe an apparent linear downward trend implying that the brighter the corona overlying a flaring source region, the less likely a high velocity CME was to occur. Furthermore, for CBI wedge values greater than approximately $1.4\times 10^{-6}$ MSB, no CMEs greater than 1,000 kms$^{-1}$ were observed. We found no events with velocity over 500 kms$^{-1}$ when the corresponding CBI value was greater than $\sim$$2.0\times 10^{-6}$ MSB. We also observed that the brighter the corona overlying a source location, the more likely for flares from that region to be non-eruptive, though we do note that only 19 flaring events were observed with CBI values greater than $\sim$$2.0 \times 10^{-6}$ MSB. However, out of these 19 events, only six were eruptive and all produced CMEs that were substantially slower than 500 kms$^{-1}$. From our results, non-eruptive flares appear to be disproportionately prevalent in regions with high overlying brightness. For example, of the ten brightest events, nine of these have no associated CME, i.e. 90\%. For the entire dataset, the rate of non-eruptive events is 53\%, which is substantially lower than the rate we see in brightest CBI events. Thus, if there is any kind of suppression present that prevents eruptions occurring, it may be limited to just the most extreme, bright events. Given these uncertainties, we do not elaborate further on eruptivity in our results, but note it as a point of interest that we will monitor as we pursue this work.

\begin{figure}[ht!]
\plotone{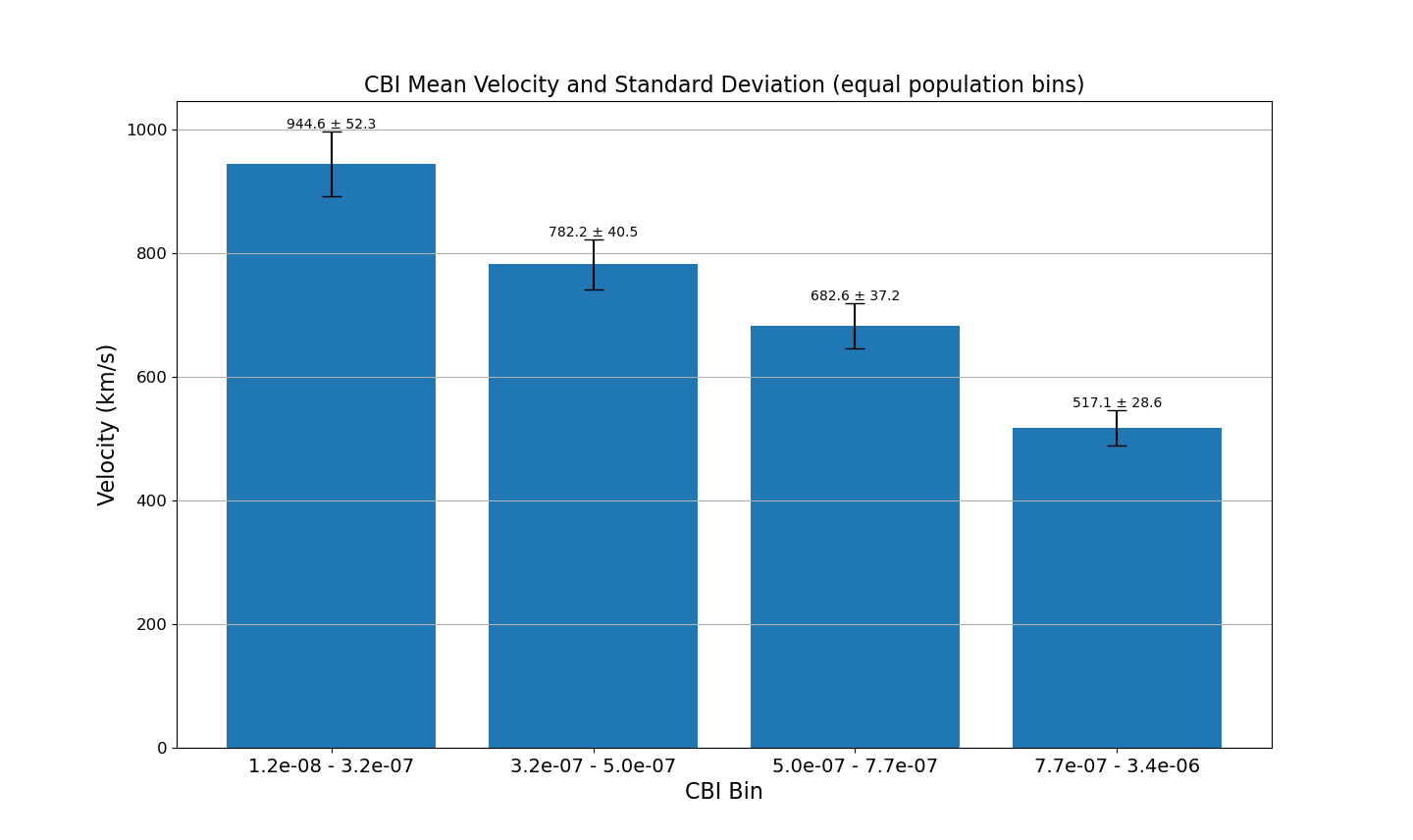}
\caption{Our results are displayed in a bar plot demonstrating an observed relationship between CME velocity in kms$^{-1}$ and CBI. The data have been split into four equal population bins according to their CBI value. Standard deviation of the mean is displayed for each bin as error bars.}
\label{fig:barplot}
\end{figure}

In Figure~\ref{fig:barplot}, we show a breakdown of the results. Here, the data have been split into four bins of equal population (i.e., 316 events in each bin) along the x-axis (CBI value). Within each bin, we compute the mean, $\mu$, and standard deviation, $\sigma$, of the velocities. For assessing whether the differences in mean velocity within the bins are statistically significant, it is the standard deviation of the mean, $\sigma_{\mu}=\sigma/\sqrt{N}$, that is more relevant, where N is the number of events per bin. Thus, in Figure 6, the error bars show $\sigma_{\mu}$. There is a decrease in mean CME velocity with CBI that looks statistically significant.

We will point out here that the anomalously bright AR12192 is responsible for all data points above $2.0 \times 10^{-6}$ MSB. As this is just one highly atypical AR, events associated with it are noted by black triangles in Figure~\ref{fig:results}. Binning the data as in Figure~\ref{fig:barplot} without the AR12192 data, still shows the same basic trend.

Later in the manuscript, we make a detailed evaluation of the possibility that our result is a statistical fluke, or an illusion, reaching a conclusion that it is most likely a real trend (see Section~\ref{sec:discuss_uncertain}, and Appendix~\ref{app:A}).

\section{Discussion} \label{sec:discuss}

\subsection{The CBI--Kinematics Connection}

The primary result from this study is an apparent relationship, or trend, between the kinematics of CMEs, and the brightness of the white-light corona overlying a given flaring source region. This is illustrated by Figures~\ref{fig:results} and~\ref{fig:barplot}, which show a significant decrease in CME velocity as localized CBI value increases. Additionally, Figure~\ref{fig:results} also shows that for summed CBI wedge values over $\sim1.4\times 10^{-6}$ MSB, no CMEs with velocity over 1,000 kms$^{-1}$ were observed to occur. Our results also highlight the atypical contribution of AR12192, which we discuss separately in Section~\ref{sec:discuss_12192} below.  

This result strongly suggests that a key process guiding the kinematics of CMEs may be directly related to, or at least signaled by, the intensity of the overlying white-light corona. While detailed analysis of any mechanisms that could potentially be causing this phenomenon are outside the scope of this initial study, there are several striking possibilities. Since the white-light that is seen in C2 is scattered free electrons, coronal brightness serves as a proxy for measuring electron density \citep[e.g.,][]{Brueckner95,Thernisien2006}. In other studies, confined flares have been associated with strong magnetic fields far from the Sun's surface \citep{Ning18}, and \cite{Singh19} found in their simulations that CME speed was inversely correlated with total pressure in the solar corona. Given that density is directly proportional to pressure, it follows that white-light coronal brightness could be a proxy for pressure as well as density. It is possible that an increase in local coronal brightness is an indicator of an increase of material in the solar corona and this material may play an important role in confining -- or at least suppressing -- eruptions. However, as we do not have in-situ measurements of this portion of the corona, and there is no scientific consensus on what causes an increase in coronal brightness, we cannot make definitive statements on causation. 

It is possible we are observing a correlation effect alone, as opposed to the electron corona being a direct causal mechanism, but nonetheless this is a potentially valuable observation. In \cite{Battams20}, it is noted that long-term time series of CBI data correlated strongly with solar irradiance measurements, and almost as strongly to open magnetic flux, again implying that stable coronal brightness is at least responsive in some manner to key coronal parameters. If CBI is indeed a reasonable proxy for pressure and density conditions in the solar corona, then white-light coronal brightness could potentially yield additional information about the magnetic field overlying a flaring source region beyond what can be evinced from disk imagery. If the mean solar brightness value alone corresponds well with these parameters, then it might be possible to correlate these features with coronal topology. Furthermore, if it is possible to learn what configurations of ARs and flaring source regions produce bright overlying coronas, then it could be possible to infer how even disk-center regions might affect CME kinematics. Again, these are speculative remarks based on these initial observations, but serve to highlight the intriguing avenues this work could advance.

\subsection{Other Statistical Observations}

\begin{figure}[ht!]
\plotone{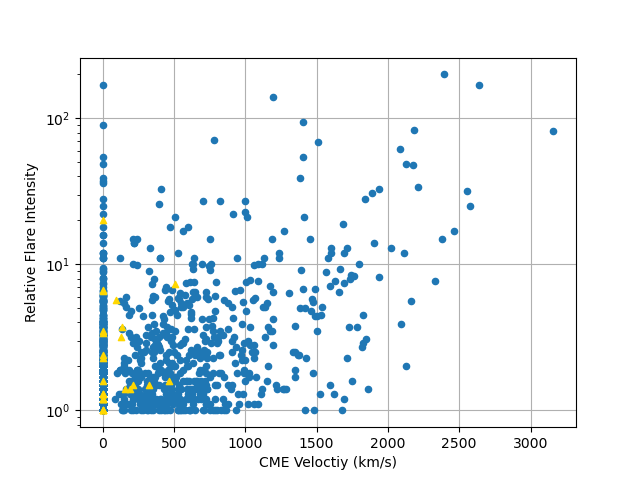}
\caption{Flare intensity plotted versus CME velocity in kms$^{-1}$ (where an M1.0-class event corresponds to an intensity of 10$^{0}$, and an X1.0 event equals 10$^{1}$). The yellow triangles represent events from AR12192, and a velocity of 0~kms$^{-1}$ means the event was non-eruptive. The blue circles are all other events in our dataset.}
\label{fig:flares_cme_vel}
\end{figure}

While the primary goal of this research is to explore the relationship between the CBI value and CME kinematics, there are a few other interesting relationships revealed by our data. The relationship between flare energies and CMEs is commonly discussed in the solar physics literature, such as by \cite{Moon03}, \cite{Chen06}, and \cite{Török07}, to name a few, and while this is not the focus of our work, we also touch on this relationship. Since we collected data by flare event, rather than by CME, we also included non-eruptive flares, thus revealing in a large statistical sample that powerful flares do not always have an associated CME. The relationship we found between flare intensity and CME is shown in Figure~\ref{fig:flares_cme_vel}, which plots the corrected CME velocity versus the relative flare intensity on a log scale. The yellow triangles are events from AR12912, and while that AR had many more X-class flares than are shown in this plot, most occurred outside the longitudinal bounds placed on our dataset. Here, the relative flare intensity is a numeric value derived from the flare class, where an M1.0 flare would receive a numeric value of 1, an M5.0 flare a value of 5, and an X1 flare a value of 10. This conversion enables the plotting of the events on a logarithmic scale, which mimics the standard logarithmic flare classification scale. Although there is a trend showing that higher velocity CMEs are more likely at higher flare intensities, there are a few high velocity CMEs (over 2000 kms$^{-1}$) at lower intensities. Similar to the results found by \cite{Yashiro05}, this suggests that the relationship between peak X-ray flux and CME velocity might not be straightforward. 

\subsection{AR12192}\label{sec:discuss_12192}

As noted in Section~\ref{sec:res}, it is worth singling out the influence of the anomalous active region 12192 on this dataset. In Figure~\ref{fig:results}, it can be seen that AR12192 dominates the brightest events in our data. Notably, all of these events produced CMEs less than 550 kms$^{-1}$, with those greater than 300 kms$^{-1}$ having lower CBI values. In other words, the overall trend of a higher CBI value being correlated with a lower velocity CME seems to persist, even when looking at just one AR. Aside from lower CME velocity, most of the events from AR12192 were entirely non-eruptive. Due to the unique nature of this AR, we do not have sufficient evidence to conclude that this is a global trend, but perhaps non-eruptivity can occur with extremely large, $\beta\gamma$ configured ARs, similar to 12192.

\subsection{Methodology Uncertainties}\label{sec:discuss_uncertain}

We arrived at these results through a novel methodology that has advantages and disadvantages. To our knowledge, white-light coronal brightness has never been used in this way, therefore this approach offers a new method of analysis for understanding the space weather environment. There are decades of white-light imagery from coronagraphs, so there is no dearth of brightness data for this type of analysis. The methodology itself is fairly straightforward and can quickly be adjusted to sample different portions of the corona with a variety of statistical approaches. The processing time for determining a CBI value, or even multiple values, is extremely short and does not rely on lengthy simulation. 

While white-light coronal brightness data is available so long as there are coronagraphs looking at the Sun, it is only the brightness on the limb of the solar disk that is visible for analysis, and any line-of-sight information is lost. Since LASCO C2 observes from L1, the occulting disk blocks the Earth-facing disk of the Sun and the line-of-sight corona is invisible. Unfortunately, this information is arguably the most critical for space weather forecasting purposes as disk-center events are the ones most likely to be associated with Earth-directed eruptions. Analyzing limb brightness offers other challenges as well, as it can be uncertain where the integrated line-of-sight brightness contributions are coming from. It is likely that some of the wedges we used to capture the brightness overlying a particular source location had contributions from other source locations, even ones behind the limb. We were careful in choosing wedges for analysis to make sure they corresponded well with the source location on the surface, but since the Sun is a complex system, it would be difficult to discretely analyze coronal brightness from one region alone. Our approach is manual, and by-eye. We are aware that this comes with uncertainties, and hope to refine this process in future work. 

Ideally, dealing with disk-center events (as viewed from Earth) would involve the use of coronagraphs operating from vantage points outside of the Sun-Earth line, e.g., L5, that would provide near-limb observations of events inside the {$\pm$}45{\textdegree} longitude bounds imposed in our study. The COR2 coronagraphs \citep{Howard08} on NASA's STEREO spacecraft\citep{Kaiser08}, provide some glimpse of this at certain times of the mission. In addition to making a CBI dataset from LASCO C2 data, we have also created one from COR2A, the coronagraph on STEREO-A, but we have not fully explored this dataset. While this will help to ameliorate the limited view we have of the solar corona, it is only useful during certain time frames since STEREO is always moving relative to Earth. We are also looking into interpolating between CBI values when a source location is near the east limb and the west limb to estimate what its value might be when the region is occulted. This could potentially provide an approximate CBI value to stand in for the unavailable data. 

As described earlier in this manuscript, the CBI data used here are fully-calibrated LASCO C2 observations, with physical units of MSB -- however, this methodology does not require them to be as such. It is simply for convenience that the CBI data exists with these units, but raw or exposure time-normalized charged-coupled device (CCD) counts, or some similar processed value, should work equally as well and may lend itself better to an operational, realtime pipeline for producing such metrics. The C2 calibration process applies linear scale factors and fixed-value offsets \citep{Battams20} to the raw C2 observations. That is, any metric for the brightness of the extracted pixels -- whether it be raw digital number (DN) values, or some other similar numerical representation -- should show the same trend, as long as the processing applied to the data applies linear transforms to pixel values (as does the process to calibrate the data to MSB values). To further this point, throughout our investigation we noted that modifying the angular width of the wedge and/or the extent (height) of the wedge into the corona, returned different mean CBI values, as would be expected. Also, individual points (events) on the plots would move around somewhat depending upon the chosen CBI wedge parameters. However, in all explorations, the global trend remained the same -- that is, the larger CBI (or equivalent metric) values routinely and consistently corresponded to slower events and, at the extreme end of the range, proportionally far more non-eruptive events. Thus, a simpler form of our presented processing pipeline would be to use raw data minus a suitable background model of some kind to remove F-coronal brightness \textemdash exactly analogous to the process described by \cite{Battams20} for creating the CBI. The advantage then would be that realtime observations could be used, thus opening up the possibility of realtime predictions (or likelihoods).

One potential concern is whether the apparent trend in Figures~\ref{fig:results} could be illusory, with the greater prevalence of high velocity CMEs at low CBI being due to there being more total CMEs at lower CBI. Caution must be exercised when interpreting apparent trends of two Gaussian-like distributions, with \cite{Wang13}, for example, noting that an apparent trend between CME speed and CME waiting time was purely a statistical correlation between two distributions. Therefore, we performed a number of tests to bolster our confidence in the strength of our results and provide statistical support that the result is most likely real. These tests took two forms: (1) a simple randomization exercise to determine whether the results shown in Figure~\ref{fig:barplot} were affected by shuffling the datasets; and (2) an attempt to create a random, statistically identical equivalent of our observations that could reproduce the trends observed in Figures~\ref{fig:results} and \ref{fig:barplot}.

The first method was to test the hypothesis that the trends we observe in our results are a consequence of some unknown phenomena that directly links the specific CBI and CME velocity values, leading to the clear trends seen in Figures~\ref{fig:results} and \ref{fig:barplot}. This premise was tested by shuffling (randomizing) the order of the individual datasets to break that linkage between data pairs. In performing this, we observed that the trend of Figure~\ref{fig:barplot} was entirely destroyed, and the trend seen in Figure~\ref{fig:results} appeared weaker. This implies that our results are likely dependent on the specific connection between CBI and CME velocity, and not based upon the underlying distributions.

The second test was based upon the premise that, assuming our results are a consequence of a statistical illusion, we should therefore be able to reliably replicate those results using purely random distributions that were statistically identical to our observations. Upon examination of histograms of the CBI and CME velocity data, we determined that distributions such as lognormal, chi-squared, and Weibull, all could be invoked to give both similar histogram presentation, as well as identical statistical properties (mean, standard deviation, range, and populations size). For all three of these distribution types, synthetic datasets were generated and their histograms inspected to ensure a strong visual representation of our observations. We then looked at both the bar plots and scatter plots for these distributions, as well as shuffled equivalents (although the latter is less important for data that is already randomly generated). In all cases, the synthetic datasets would return scatter plots that somewhat resembled Figure~\ref{fig:results}, albeit not as ``clean'' as our observations, but none returned bar plots that showed the clear trend observed in Figure~\ref{fig:barplot}. This test demonstrates that our results cannot be fully replicated using near-identical statistical distributions of random data, thus reinforcing the idea that the specific connection between the CBI and CME velocity observations is crucial to the clear trends we observe. A complete description of these statistical tests, including sample plots, is provided in Appendix~\ref{app:A}. 

While it remains impossible to prove with absolute certainty that our trend is not an exceedingly rare statistical fluke, our statistical tests repeatedly failed to adequately replicate our results, and randomization of our data largely destroyed the observed trends. Thus, we are confident that some direct or indirect connection does exist between coronal brightness and CME velocity, and indicates a relationship that is worthy of more detailed examination in follow-up studies.

Finally, the results we present here, even without detailed knowledge of the physical mechanism, may be promising to the space weather forecasting community. The CBI dataset and this method of analysis demonstrate that conditions of the white-light solar corona as seen in LASCO C2, might be important indicators of space weather drivers. We have shown that CBI is a potential predictor of CME velocity. CME velocity has, in turn, been correlated with geoeffectivenes \citep{Yurchyshyn04}, as measured by various geomagnetic indices such as the magnitude of the southward magnetic field and the disturbance storm time \citep[e.g.,][]{Besliu-Ionescu21, Vourlidas19}. CME velocity is also correlated with solar energetic particle (SEP) generation, which can originate from CME shocks \citep[e.g.,][]{Frassati22, Gopalswamy04, Kahler01}, and it is these SEPs that are potential threats to astronauts journeying beyond the magnetosphere to the Moon or Mars \citep[e.g.,][]{Guo23}. Currently, we only know when a CME will occur once it has already left the Sun and is visible in a coronagraph. This method, if proven repeatable in future analyses, could potentially increase the time for space weather forecasters to know if a high-velocity CME is likely to occur or not. Efforts are already underway to produce a simplified CBI data product and software tool to be tested in a realtime setting. 

\section{Conclusion} \label{sec:conc}

We present a novel analysis of white-light coronal brightness and its relationship to the velocity of flare-associated CMEs. We found the following:

\begin{itemize}
\item The brighter the corona overlying a flare source location, the less likely a high-velocity CME is to occur.
\item For CBI wedge values greater than approximately $1.4\times 10^{-6}$ MSB, no CMEs faster than 1,000 kms$^{-1}$ were observed to occur. 
\item Active Region 12192 is associated with the 20 brightest events in our data. While this AR is anomalously bright, it demonstrates the same trend found in the rest of our dataset, the brighter the overlying corona is, the lower the velocity of the associated CMEs. Even without the AR12192 events, we still found that the brighter the overlying coronal field, the lower the mean CME velocity.
\item As a by-product of this study, we observe the established trend that CME velocity correlates with flare intensity.
\end{itemize}

Though determining the physical mechanism responsible for this apparent trend is outside the scope of this paper, one of the possible causes of higher values of coronal brightness is increased magnetic field strength overlying a flaring source region. AR12192, which dominates our results, has a strong overlying magnetic field \citep{Sun15} and it is the brightest event from our dataset. It is possible a stronger overlying magnetic field could cause an increase in pressure thus inhibiting CME velocity \citep{Sarkar18}. It is much harder to speculate on CME eruptivity as a function of overlying coronal brightness, as these two processes may be distinct. But the result is nonetheless one of interest. Further research will have to be done to determine any causal or correlational mechanisms, but these are promising lines of inquiry. 

Our result can potentially enhance space weather forecasts, with advanced knowledge of CME kinematics to be expected from a region on the solar surface. Currently, work is being done to develop a realtime version of the coronal brightness index, that will help to test the value of using CBI as a forecasting metric. 




\begin{acknowledgments}
Financial support was provided by the Office of Naval Research, and by NASA award 80HQTR21T0006 to the Naval Research Laboratory. The authors would like to thank Dr. Seiji Yashiro for providing flare-associated CME data. 
\end{acknowledgments}

%






\appendix

\section{Statistical Testing}\label{app:A}

As noted in Section~\ref{sec:res}, care must be taken when interpreting apparent trends in scatter plots of two Gaussian-like distributions. Sometimes such trends are in fact illusory, as demonstrated by Figure. 4 in \cite{Wang13}, which showed that an apparent trend between CME speed and CME waiting time was purely a statistical correlation between two distributions. In this appendix, we present the additional statistical tests we performed to gain confidence in the legitimacy of our reported relationship between CBI and CME velocity (``Vcme''). 

While it is not possible to demonstrate beyond reasonable doubt as to whether a result is coincidence or not, we can perform a variety of statistical tests to gain confidence in the interpretation of our result. Therefore, we took two approaches to testing the statistical strength of our results. The first approach was based upon randomization of the observations in an attempt to break the connection between individual data points. The second approach was to produce synthetic datasets for both CBI and CME velocity that were statistically identical (size, mean, standard deviation, and range) to our dataset, and then apply the same analyses to these datasets to see if the conclusions were the same.

\subsection{Randomization}

The first test was based upon exploring the premise that the trend observed in our results is a consequence of some unknown phenomenon that directly links the specific CBI and CME velocity values, leading to the clear trend seen in Figures~\ref{fig:results} and \ref{fig:barplot}. If this result is an illusion, we should be able to randomize the order of the individual data points, thus breaking the individual x-y connections, and scatter and bar plots (i.e., Figures~\ref{fig:results} and \ref{fig:barplot}) should retain the same trend. If, however, the trends are diminished or gone, then we can be more confident that the result likely relies on the specific connection between the individual x-y (CBI--Vcme) data points.

\begin{figure}[ht!]
\plotone{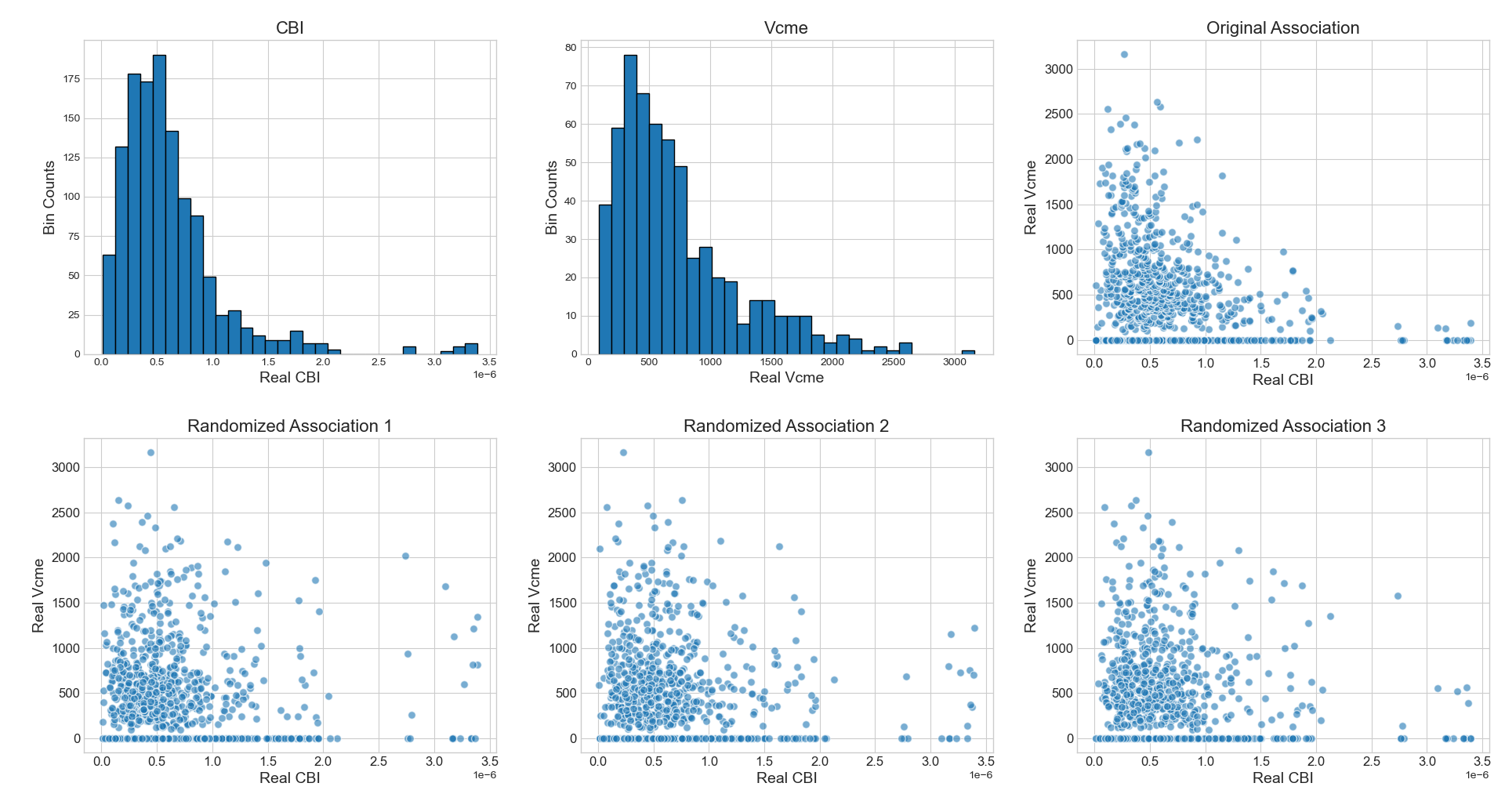}
\caption{Panel a shows a histogram of the CBI data, panel b shows a histogram of the velocity data, panel c is the original association between CBI and Vcme, and panels d, e, and f, are randomizations of both variables.}
\label{fig:real_cbi}
\end{figure}

\begin{figure}[ht!]
\plotone{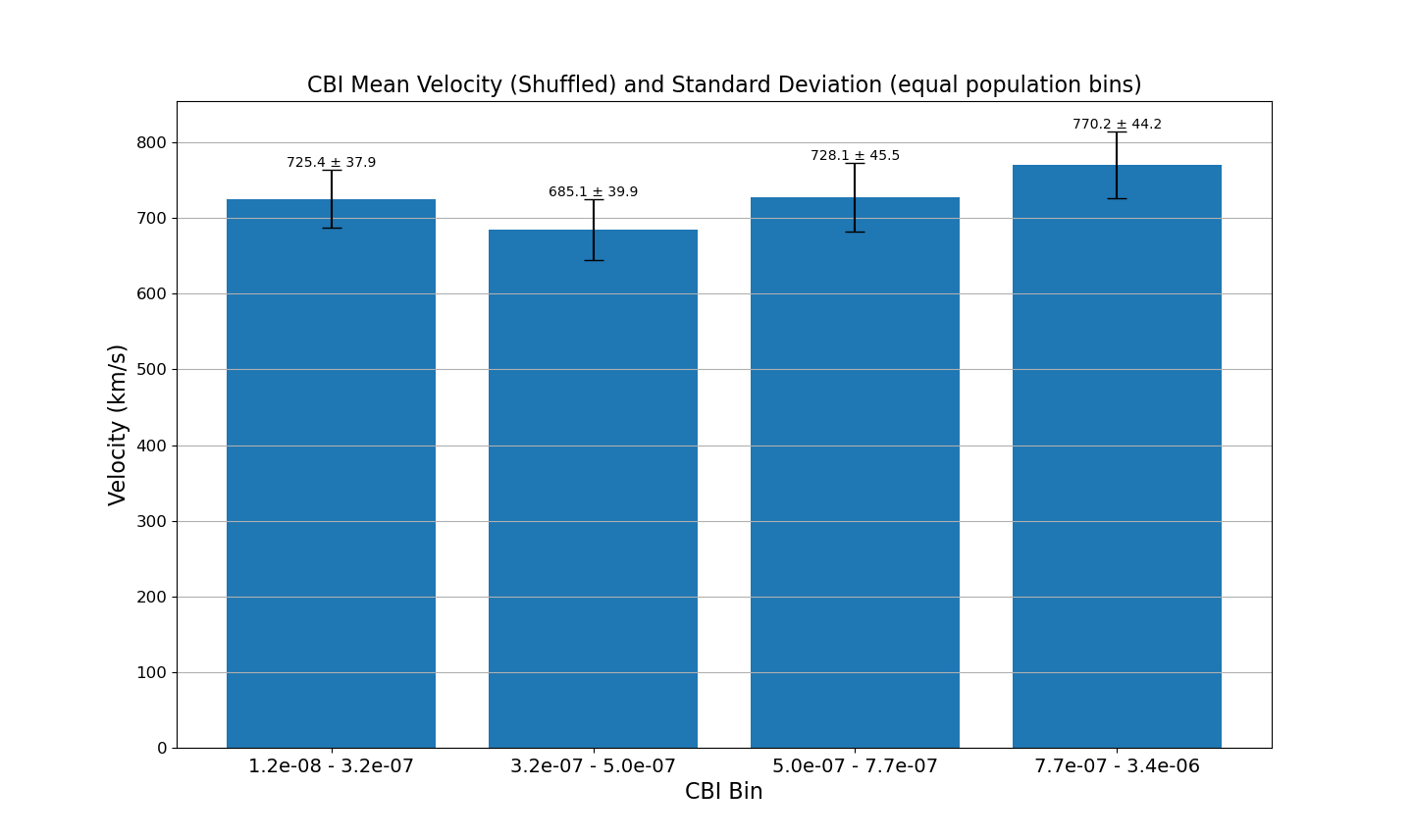}
\caption{The real data has been shuffled along both variables. Standard deviation of the mean is displayed for each bin as error bars. }
\label{fig:shuffbar}
\end{figure}

Figure~\ref{fig:real_cbi} shows the histograms for the CBI (panel a), and Vcme (panel b). We also show the original association (panel c). The second row (panels d, e, f) shows scatter plots after randomizing the order of both CBI and Vcme. In practice, we repeated this many times, but only show three examples here. While it appears in the randomized plots that there is, arguably, still a similar trend to the original trend (albeit weaker), the corresponding bar plots for these distributions show that the trend has indeed been destroyed. This is seen in Figure~\ref{fig:shuffbar}, which shows a shuffling of the original association Figure~\ref{fig:real_cbi}(c) of both CBI and Vcme. Once again, we performed this shuffling many times, but only included one result here. This result should be compared to Figure~\ref{fig:barplot}, which shows a clear trend. Thus, randomization of our data appears to be extremely detrimental to our results.

\subsection{Synthetic Data Exploration}

Our second test sought to employ three different distribution types -- lognormal, chi-squared, and Weibull -- to generate synthetic datasets that statistically matched our datasets (including visual similarity to histograms) to explore the extent to which our result could be replicated using only random data. The hypothesis was that, should our results be a statistical fluke, similar to that reported by \cite{Wang13}, they should be largely replicable by random value distributions. In all three of the following examples, the distributions were generated using statistics identical to that of our observations, namely population size, mean, standard deviation, and range. We note that some trial and error was required for the various algorithmic parameters in order to best replicate the shape of the observed distributions.

\begin{figure}[ht!]
\plotone{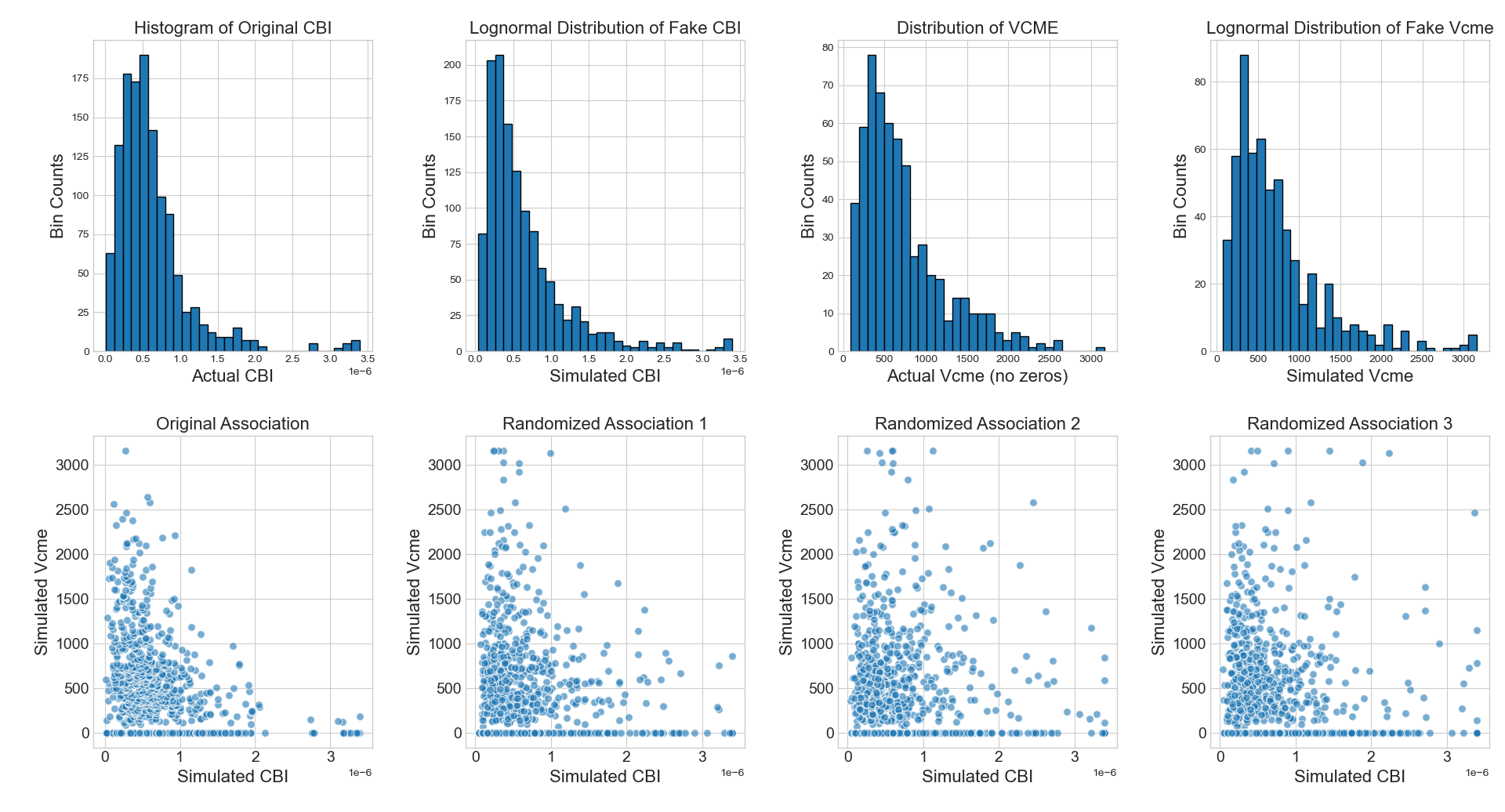}
\caption{This lognormal distribution was simulated using the statistics from our data. For convenience, panel a shows a histogram of the original CBI data, panel b shows a histogram of the simulated, lognormal CBI data. Panel c shows a histogram of the original CME velocity data, and panel d shows a histogram of the simulated, lognormal, Vcme data. Panel e is the original association between the simulated lognormal, datasets, and panels f, g, and h, are scatter plots were both variables have been randomized.}
\label{fig:lognorm_stat}
\end{figure}

\begin{figure}[ht!]
\plotone{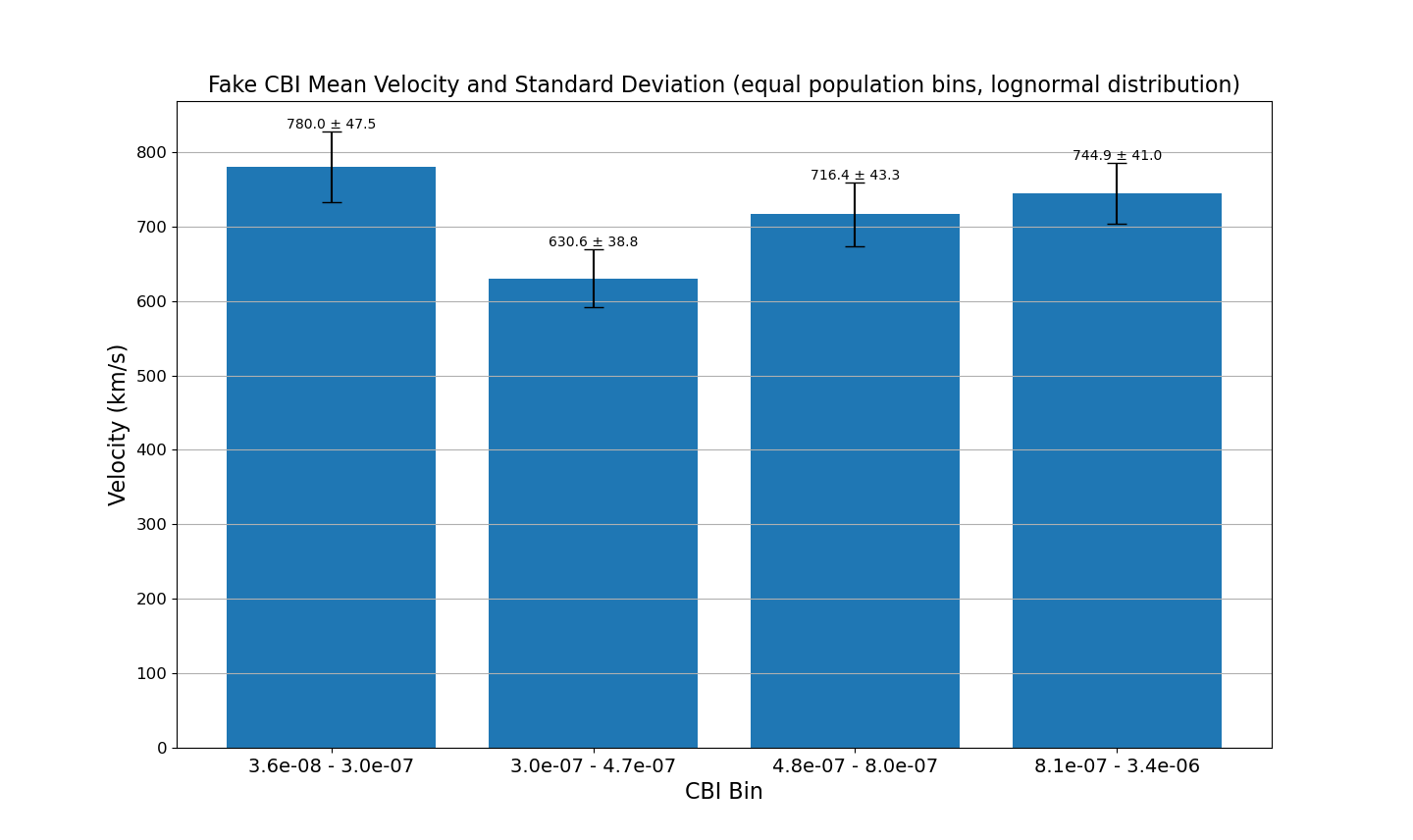}
\caption{Simulated data with a lognormal distribution is placed in four bins of equal size, similar to the process for Figure~\ref{fig:barplot}. Standard deviation of the mean is displayed for each bin as error bars.}
\label{fig:logbar}
\end{figure}

First we looked at a lognormal distribution, with the results presented in Figure~\ref{fig:lognorm_stat}. The layout of this figure is the same as Figure~\ref{fig:real_cbi}, with the addition of histograms of our real data next to the simulated data, for comparison. Similar to when we randomized our actual observations, the scatter plots look comparable to the original association, showing an apparent downward trend in velocity as CBI values increase, though the association is arguably weaker and some of the right skewness has been lost. Repeating this test many times (not shown) typically does result in a scatter plot with a downward trend, similar to our original result, and shuffling the synthetic observations makes no substantial difference, as expected. However, despite this apparent visual similarity to our observations, the bar plot that results from this lognormal simulation does not show the same linear downward trend as Figure~\ref{fig:barplot}. An example bar plot, with the data from Figure~\ref{fig:lognorm_stat}(e) placed into four bins of equal population size is shown in Figure~\ref{fig:logbar}. Since this simulated data is random, this bar plot is broadly representative of all simulated, lognormal distributions. This is a key point: the apparent trend that is visually observed in the very dense scatter plots is not necessarily reflective of the true shape of the distribution, which is better summarized by bar plots of equal bin population.

\begin{figure}[ht!]
\plotone{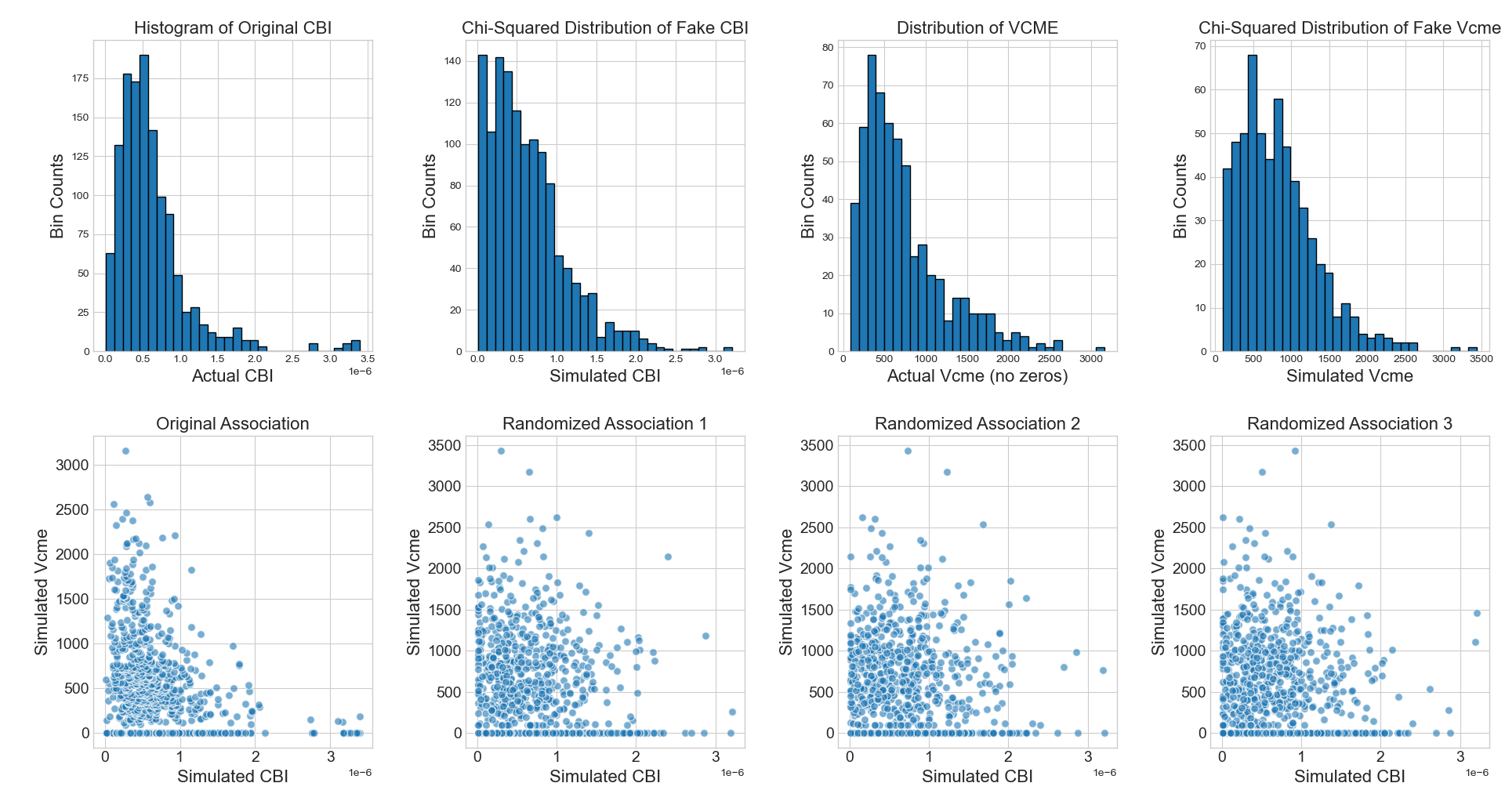}
\caption{Simulated data was randomly generated using a chi-squared distribution, with statistics identical to that of our observations. The layout of this figure is the same as Figure~\ref{fig:lognorm_stat}.}
\label{fig:chsq}
\end{figure}

\begin{figure}[ht!]
\plotone{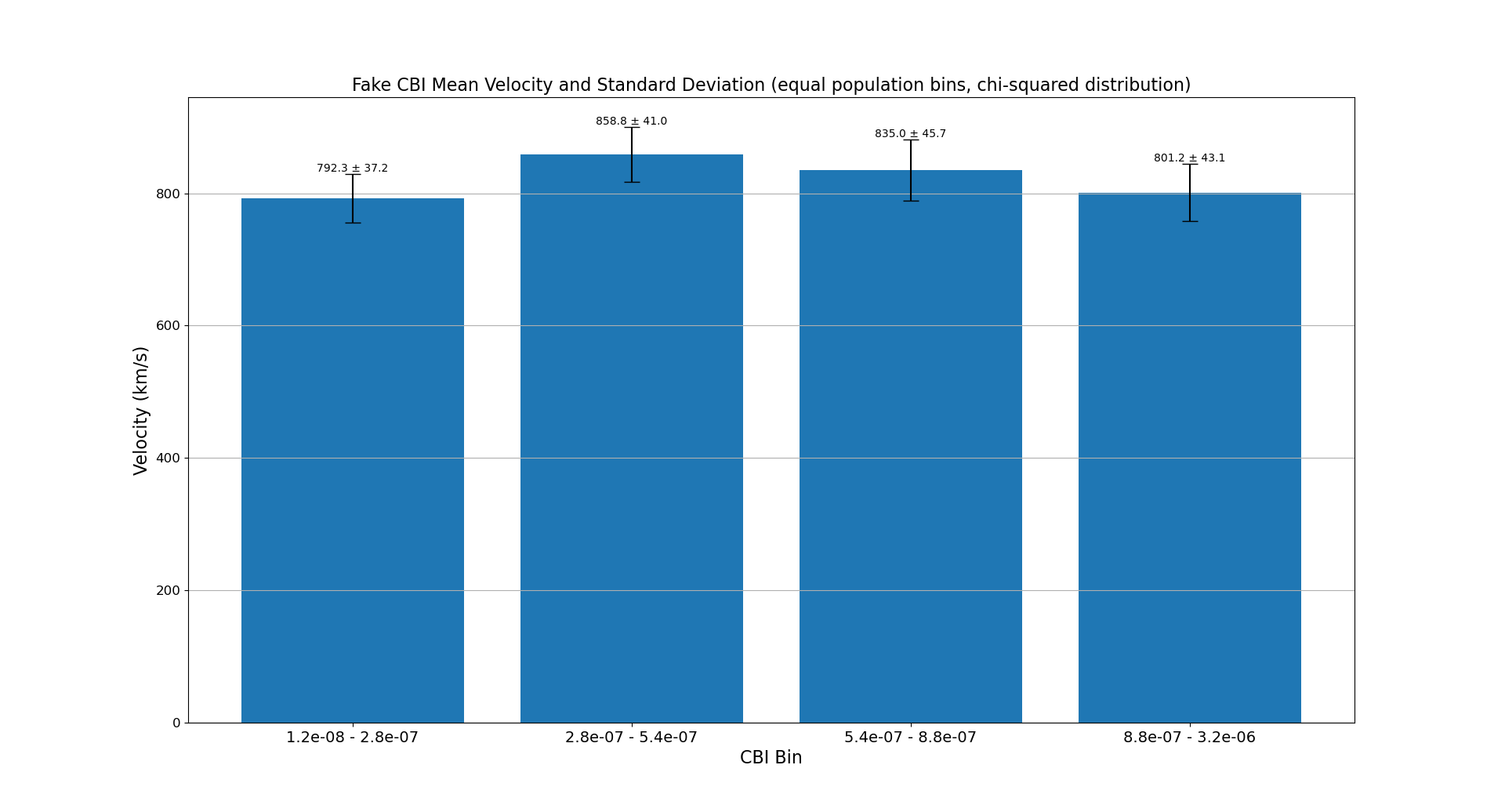}
\caption{Simulated data was randomly generated using a chi-squared distribution, with statistics identical to that of our observations. The layout of this figure is the same as Figure~\ref{fig:logbar}.}
\label{fig:chsqbar}
\end{figure}

Next, we looked at a simulated chi-squared distribution. The results are presented in Figure~\ref{fig:chsq}. This figure has an identical layout to that of Figure~\ref{fig:lognorm_stat}. Again, we find that the scatter plots are similar to those of Figure~\ref{fig:real_cbi}. This is true of both the original association, Figure~\ref{fig:chsq}(e), and the randomized associations of Figure~\ref{fig:chsq}(f-h). The randomized scatter plots, however, lose some of the downward trend of the original association. Similar to the situation we found with the lognormal distribution, the apparent visual similarity to our original dataset is non-existent in bar plots of the chi-squared simulations. A bar plot of the data in Figure~\ref{fig:chsq}(e) can be found in Figure~\ref{fig:chsqbar}. Since the simulated data is random, this bar plot is broadly representative of all simulated, chi-squared distributions. Once the cluttered scatter plot is placed in bins of equal population, it can be seen that the clear downward trend of our real dataset is non-existent in the chi-squared simulation.

\begin{figure}[ht!]
\plotone{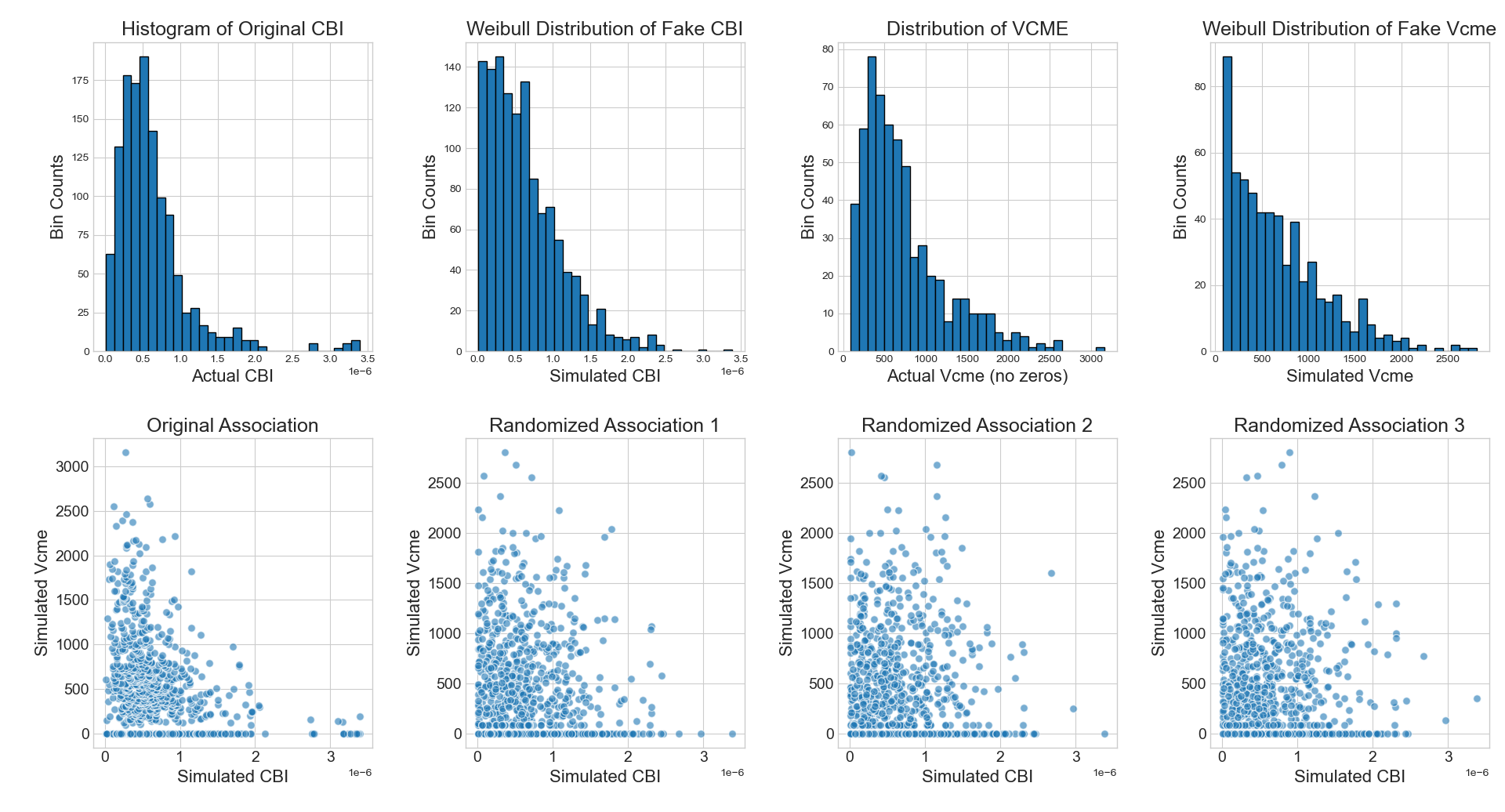}
\caption{Simulated data was randomly generated using a Weibull distribution, with statistics identical to that of our observations. The layout of this figure is the same as Figure~\ref{fig:lognorm_stat}.}
\label{fig:wb}
\end{figure}

\begin{figure}[ht!]
\plotone{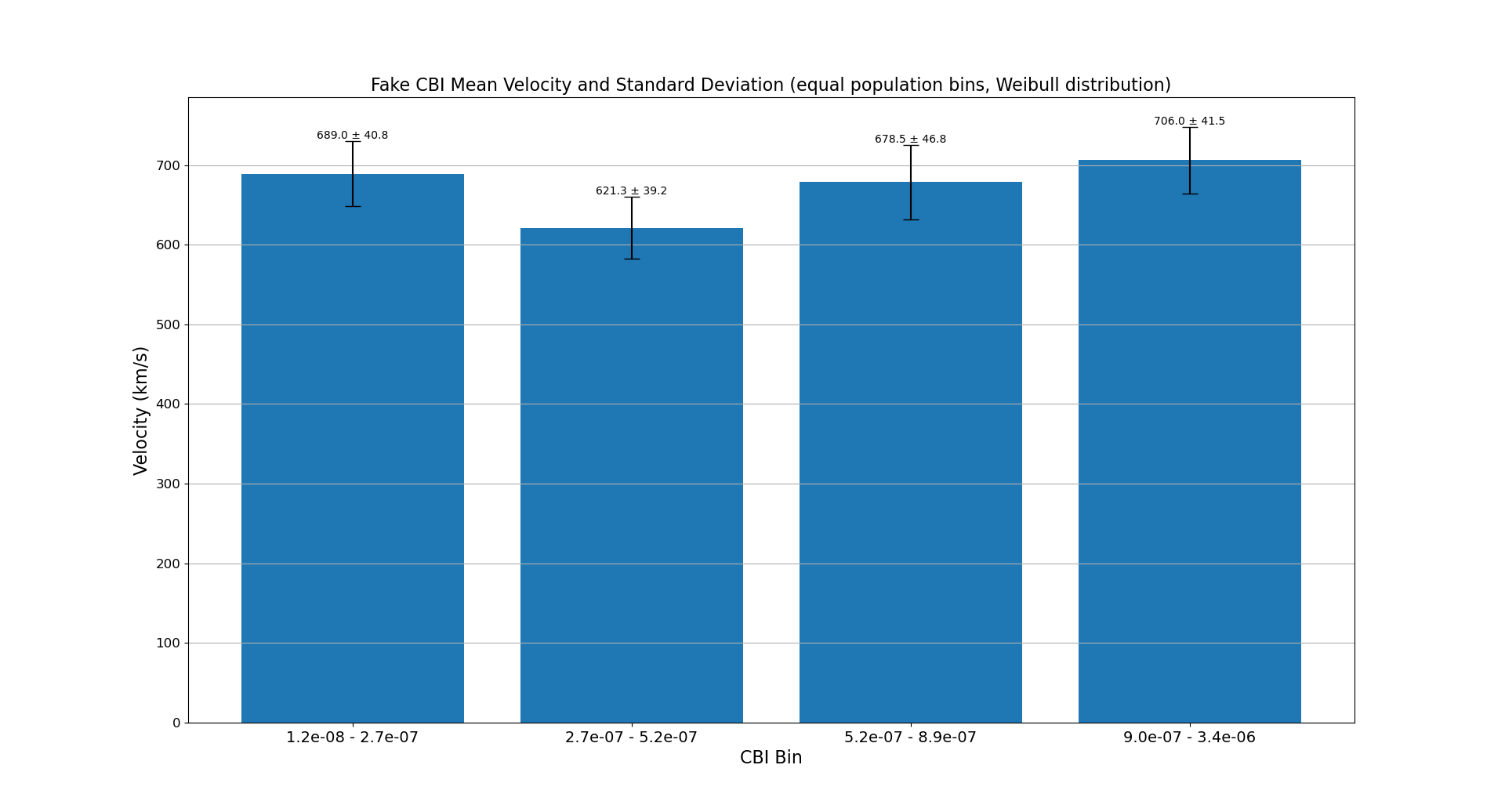}
\caption{Simulated data was randomly generated using a Weibull distribution, with statistics identical to that of our observations. The layout of this figure is the same as Figure~\ref{fig:logbar}.}
\label{fig:wbbar}
\end{figure}

Lastly, we explored a synthetic dataset with a Weibull distribution, as shown in Figure~\ref{fig:wb}. The data has been displayed in the same format as the lognormal and chi-squared distributions. Like the previous simulated datasets, we again find similarity to the shape of the distribution as seen in the histograms and scatter plot of our original datasets. Once the data is randomized, some of the apparent right skewness is lost. Once again, as we demonstrated with the previous simulations, the similarity to the CBI-Vcme trend ends when we do a deeper investigation into the dense data of the scatter plots. Figure~\ref{fig:wbbar} shows a bar plot of the Weibull distribution data found in Figure~\ref{fig:wb}(e), but since this data is random, this bar plot is representative of all these simulated Weibull distributions. In this format, we can see that the shape of the distribution is unlike that of our original CBI and CME velocity datasets.

These statistical tests demonstrate that while the apparent trend we observe in our scatter plot, Figure~\ref{fig:results}, is somewhat reproducible via a number of random distributions, the underlying trend in the observations -- as shown in Figure~~\ref{fig:barplot} -- cannot be replicated by these synthetic distributions. These tests also highlight the importance of appropriate binning of data for such visualizations, versus relying on apparent trends in visually dense scatter plots.

\clearpage 

\bibliography{bibliography}{}
\bibliographystyle{aasjournal}



\end{document}